\documentclass[twocolumn,showpacs,longbibliography,nofootinbib,pra]{revtex4-1}
\usepackage{graphicx}
\usepackage{textcomp}
\usepackage{amsmath,amssymb}
\usepackage{color}
\usepackage{mathtools}

\usepackage{cleveref}
\crefname{figure}{Fig.}{Fig.}
\Crefname{figure}{Fig.}{Fig.}
\crefname{equation}{eq.}{eqs.}
\Crefname{equation}{Eq.}{Eqs.}

\DeclarePairedDelimiter\st{|}{\rangle}
\DeclarePairedDelimiter\floor{\lfloor}{\rfloor}

\begin{document}

\title{Controlled generation of genuine multipartite entanglement in Floquet Ising spin models}
\author{Gautam Kamalakar Naik}
\email{gautamk.naik.phy15@itbhu.ac.in}
\affiliation{ Department of Physics, Indian Institute of Technology (Banaras Hindu University), Varanasi - 221005, India}
\author{Rajeev Singh}
\email{rajeevs.phy@itbhu.ac.in}
\affiliation{ Department of Physics, Indian Institute of Technology (Banaras Hindu University), Varanasi - 221005, India}
\author{Sunil Kumar Mishra}
\email{sunilkm.app@iitbhu.ac.in}
\affiliation{ Department of Physics, Indian Institute of Technology (Banaras Hindu University), Varanasi - 221005, India}
\date{\today}

\begin{abstract}
We propose a method for generation of genuine multipartite entangled states in a short-range Ising spin chain with periodic global pulses of magnetic field.
We consider an integrable and a non-integrable Floquet system that are periodic in time and have constant quasi-energy gaps with degeneracies. We start with all spins polarized along one direction and show that they evolve into states with high entanglement by calculating the average entanglement entropy and geometric measure of entanglement. We show that some of these states have a high number of parties involved in the entanglement by calculating the quantum Fisher information. Such controlled generation of multipartite entanglement has potential applications in quantum information processing.
\end{abstract}


\maketitle

\section{Introduction}
Genuine quantum correlations as encapsulated by quantum entanglement give rise to effects which have no counterparts in classical physics \cite{Amico2008a,Horodecki2009}. Quantum entanglement acts as a resource in quantum information science and is exploited for quantum teleportation \cite{Zeilinger2018}, quantum computation \cite{Gottesman1999,Bennett2000,Ekert1996} and quantum cryptography \cite{Kempe1999,Jennewein2000a,Naik2000,Tittel2000}. 
Entanglement leads to understanding the fundamentals of various phases in many-body systems and detection of quantum phase transitions \cite{Osterloh2002,Singh2016,Ponte2014,Kj??ll2014}.
The notion of multiparty entanglement, i.e. entanglement when the system is composed of more than two subsystems, is not straightforward and has been a field of research by itself \cite{Guhne2006,Guhne2005,Brub2005,Stelmachovic2004}. Classifications and various interpretations of multipartite entanglement have been actively discussed in the past \cite{Szalay2015,Yu2014,Gour2013,Huber2010}. Multipartite entanglement has proven to be a valuable resource for quantum computation and information \cite{Kempe1999,Yeo2006}.
The geometric measure of entanglement, which is one of the many multipartite entanglement quantifiers, does not explicitly consider subsystems and measures the overall entanglement in the system \cite{Wei2003a,Das2016,Blasone2008}. However, by itself, it fails to give information about the number of parties involved in the entanglement.
Another measure called the quantum Fisher information, which is an indisposable part of modern quantum metrology, gives a lower bound on the number of parties involved in the entanglement \cite{Hauke2016,Toth2012,Hyllus2012,Pezze,Mirkhalaf2017}.


The physical implementations of quantum information technologies is carried out in cold atoms \cite{Bloch2012}, optical systems \cite{Miller2005} and various condensed matter systems such as quantum spins, superconducting qubits and quantum dots \cite{Gershenfeld1997,Nakamura1999,Walther2006}.
For a large class of physical systems, there exist effective spins models that describe the relevant processes \cite{Smith,Yao2014}. 
The transverse field Ising model, which is one of the paradigmatic models of quantum phase transitions \cite{Heyl2013,Cole2017}, is exactly solvable using Jordan-Wigner transformations \cite{Lieb1961,Pfeuty}. But upon addition of longitudinal magnetic field it becomes non-integrable as the Jordan-Wigner transformation in this case gives an interfermionic interaction term. 
The analytical study of the Ising Floquet system has been done for the integrable cases by \textcite{Prosen2004,Prosen2000,Prosen2002,Lakshminarayan2004,Else2016}.

In this article, we study the integrable periodically kicked transverse field Ising model and the non-integrable model with an additional longitudinal field, both with a specific driving period.
We focus on the entanglement structure during the time evolution of simple product initial states. We show that these Floquet systems are periodic in time by studying their quasi-energies. We also show that the time evolved states have high multiparty entanglement by calculating quantifiers such as average entanglement entropy and the geometric measure of entanglement. We calculate the quantum Fisher information of the time evolved states to get the lower bound of the number of parties involved in the entanglement.
These measures help us identify states with high genuine multiparty entanglement that are obtained during the time evolution (such as the GHZ states). 
We note that a method to prepare GHZ states and W-states in a long range Ising spin chain has been recently proposed \cite{Chen2017}. But we would like to point out that 
our scheme uses a system with only short range couplings which may be easier to control.

This article is organized as follows. In section \ref{modelsec} we discuss the Floquet map of the system. Subsequently, in section \ref{periodicity} we describe the the periodic nature of the system. In section \ref{multiparty-sharing} we present the results of numerical calculations of the average entanglement entropy, geometric measure of entanglement and quantum Fisher information and the implications of these results to the nature of entanglement seen in these systems. We summarize the main results and point out the future directions in section \ref{conclusion}. 

\section{Model}
\label{modelsec}
In this work we consider a periodically driven Ising system  (which will be referred as
the $\mathcal{U}_x$ system hereafter) whose dynamics is given by the Floquet operator (which is the time evolution operator over one period)
\begin{equation}
\label{U_x}
\mathcal{U}_x=\exp\left(-i\frac{\pi}{4} (\mathcal{H}_{xx}+\mathcal{H}_{x}) \right)  \exp\left(-i\frac{\pi}{4}H_{y} \right),
\end{equation}
where $\mathcal{H}_{xx}=\sum_{i=1}^{L-1} {\sigma_i^x \sigma_{i+1}^x}$ is the
nearest neighbor Ising interaction term with unit interaction strength, 
$\mathcal{H}_{x}=\sum_{i=1}^L {\sigma_i^x}$ is the longitudinal field term  and $\mathcal{H}_{y}=\sum_{i=1}^L {\sigma_i^y}$ is the transverse field
in $y$-direction. 
A system with Hamiltonian given by
\begin{equation}
\label{H_t}
\mathcal{H}(t)= \mathcal{H}_{xx} +\mathcal{H}_{x}+ \sum_{k= - \infty }^{\infty} \delta \left( \frac{t}{\pi/4}-k \right) \mathcal{H}_{y} ,
\end{equation}
where the transverse magnetic field in $y$-direction is applied at a regular interval of $\pi/4$ time period in the form of delta pulses, would have states just before application of consecutive delta pulses related by the unitary Floquet map (as in \cite{Berry1979}) given in \cref{U_x}.
The presence of the longitudinal term in the Hamiltonian ceases the possibility of finding the exact solution using the Jordan-Wigner transformation.
Therefore, we will explore the numerical solutions of this model using exact diagonalization.
The Floquet operator in \cref{U_x} is equivalent to the Floquet operator
\begin{equation}
\mathcal{U}_x=\exp\left(-i\frac{\pi}{4}\mathcal{H}_{xx} \right) \exp\left(-i\frac{\pi}{4}\mathcal{H}_{z} \right) \exp\left(-i\frac{\pi}{4}\mathcal{H}_{x} \right).
\end{equation}


We shall also consider an integrable model, termed as $\mathcal{U}_0$ system, with the Floquet operator given by
\begin{equation}
\label{U_0}
\mathcal{U}_0=\exp\left(-i\frac{\pi}{4}\mathcal{H}_{xx} \right) \exp\left(-i\frac{\pi}{4}\mathcal{H}_{z} \right),  
\end{equation}
where $\mathcal{H}_{xx}=\sum_{i=1}^{L-1} {\sigma_i^x \sigma_{i+1}^x}$ and $\mathcal{H}_{z}=\sum_{i=1}^L {\sigma_i^z}$ and open boundary 
conditions are considered. 
The above model has been shown to be useful in generating states with multiple Bell pairs \cite{Mishra2015}.


In this article we study the entanglement structure in the $\mathcal{U}_0$ and $ \mathcal{U}_x$ systems for different initial states. Hereinafter, we refer to the states with all individual spins being eigenstates of $\sigma^\alpha$ as $\alpha$-states and initial states with all individual spins being eigenstates of $\sigma^\alpha$ as $\alpha$-initial states, where $\alpha \in \{x,y,z\}$.
In the subsequent sections we will analyze the systems with even system sizes having open boundary conditions unless stated otherwise. We comment on the other cases (with different boundary conditions and system sizes) in the Appendix \ref{summary_table}.

\section{Periodicity}
\label{periodicity}

\begin{figure}
	\centering
	\includegraphics [angle=0,width=\linewidth] {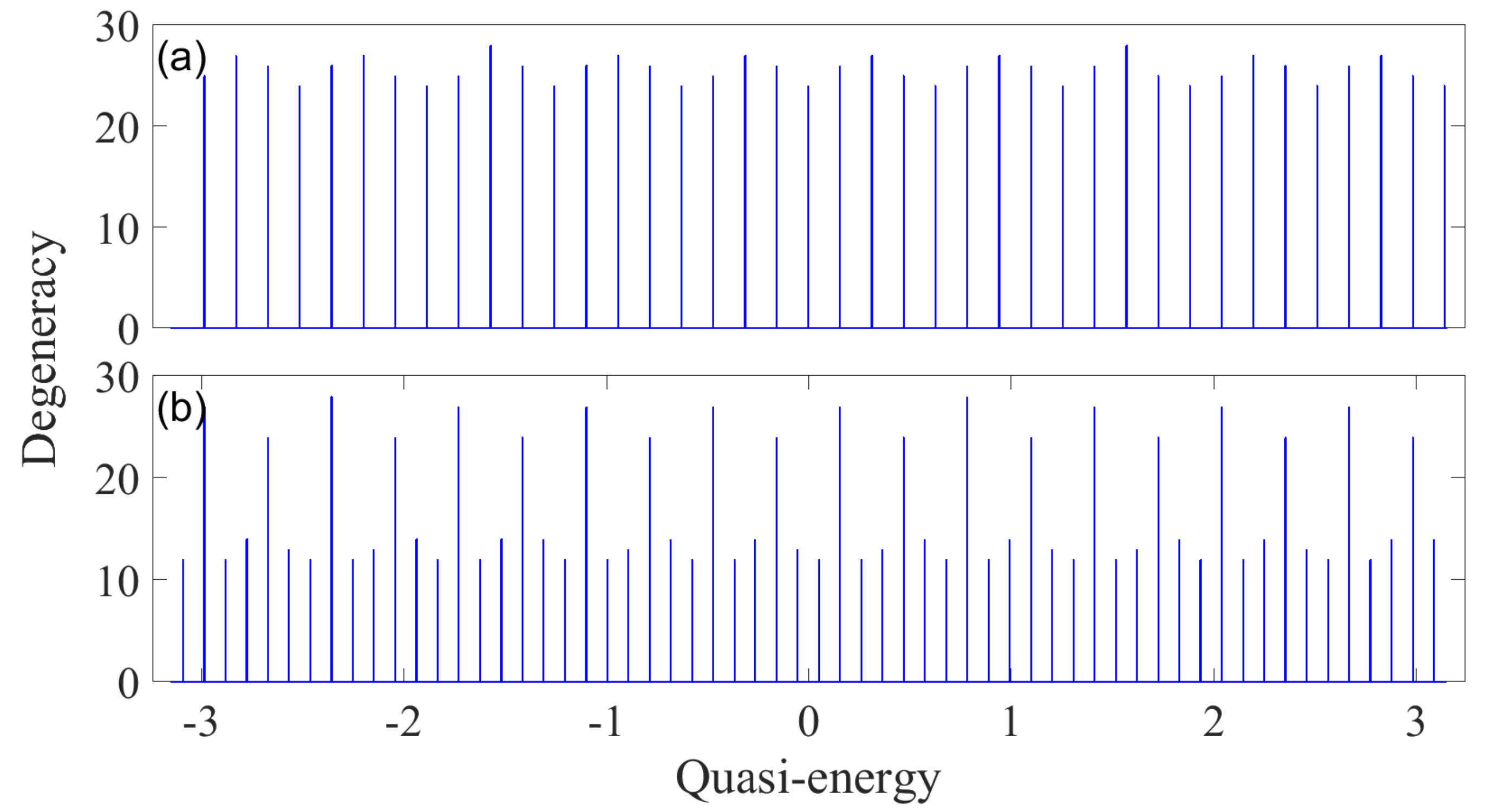}
	\caption{Degeneracy of quasi-energies in the $\mathcal{U}_0$ (a) and the $\mathcal{U}_x$(b)  systems for system size $L=10$. Both systems have equally spaced quasi-energies. The $\mathcal{U}_0$ system has spacings of $\pi/(2L)$. The spacings in the $\mathcal{U}_x$ system is not a simple function of the system size. }	
	\label{degeneracy}
\end{figure}

There is periodicity in the entanglement profile and it is the outcome of constant quasi-energy gaps of the Floquet system. The eigenvalues of the Floquet operator are of the form $e^{-i \theta_k}$ where $ \theta_k \in [-\pi,\pi]$ are known as the quasi-energies of the Floquet system. The degeneracy of the quasi-energies of the system is shown in \cref{degeneracy}. The $\mathcal{U}_0$ system with open boundary conditions has quasi-energies in the multiples of $\pi/ (2L)$ and hence the system is periodic with period $2L$ (as $\mathcal{U}_0^{2L}=I$). The $\mathcal{U}_x$ system, however, does not have a simple relation for the quasi-energies in terms of the system size but the quasi-energies are still equally spaced for a given system size. For example, a system size of $L=10$ has quasi-energies which are odd multiples of $\pi/60$.

\section{Probing Multipartite Entanglement}
\label{multiparty-sharing}

We study two aspects of the multipartite entanglement in the systems under consideration, the extent of entanglement between the different particles of the system and the number of particles that get entangled. Measuring these aspects of multipartite entanglement are research areas in their own respect and there are several quantifiers for the above purposes, each with their own advantages and shortcomings. Here we study the extent of entanglement through the average entanglement entropy (AEE) and the geometric measure of entanglement. We calculate the quantum Fisher information (QFI) to get a lower bound on the number of particles that are involved in the entanglement.

\begin{figure}[t!]
\centering
\includegraphics [angle=0,width=\linewidth]{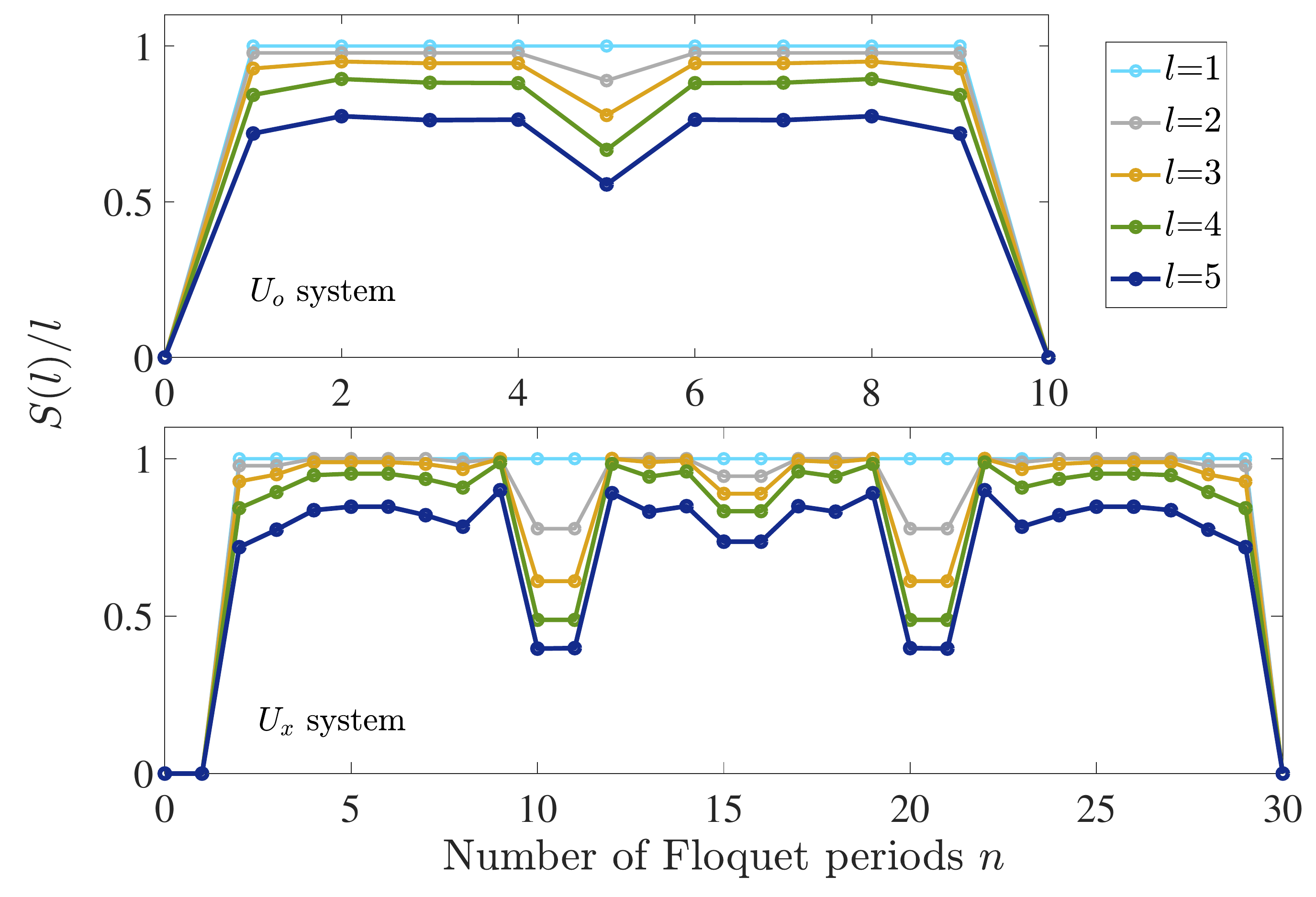}
\caption{Variation of the normalized average entanglement entropy (AEE) $S(l)/l$ for the $\mathcal{U}_0$ and $\mathcal{U}_x$ system with successive Floquet periods in a system size $L=10$ with the $z$-initial state.}
\label{n10En}
\end{figure}

The average entanglement entropy (AEE) of the system over all the partitions of subsystem size $l$ is given by:
\begin{equation}
S(l)= \langle -\text{Tr} \big( \rho_{P_l} \log(\rho_{P_l}) \big) \rangle_{P_l} ,
\label{S_k}
\end{equation}
where $\rho_{P_l}$ represents the reduced density matrix of a partition $P_l$ of size $l$ and $\langle \cdot \rangle_{P_l}$ represents the average over all possible partitions $P_l$. The AEE plots in \cref{n10En} suggest that the average entanglement is high throughout the periodic cycle and is spread out among the subsystems. A few points in the $\mathcal{U}_x$ system where we find a drop in the AEE values are those points corresponding to states that have large number of particles involved in the entanglement (as will be shown in the next section). The $\mathcal{U}_0$ system comes back to its initial state after the 10th Floquet period while the $\mathcal{U}_x$ system reaches its flipped state after the 30th Floquet period (\cref{n10En}). Hence the entanglement profile repeats itself beyond these points.

%

\begin{figure}[t]
\centering
\includegraphics[angle=0,width=\linewidth]{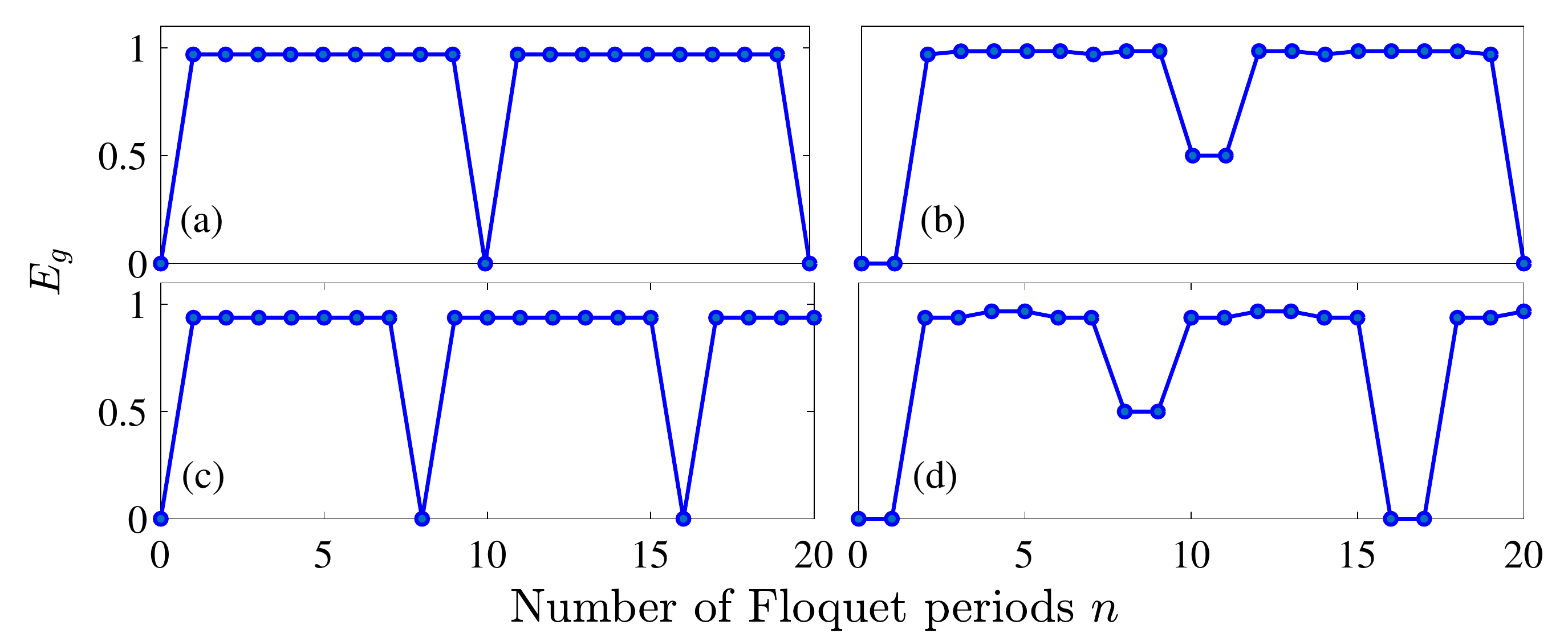}
\caption{Variation of the geometric measure of entanglement $E_g$ with consecutive Floquet periods for the $\mathcal{U}_0$ system sizes $L=10$ (a, b) and $L=8$ (c, d). In the case of the $z$-initial state (a, c), we see here that this entanglement measure does not distinguish between the state with $L/2$ Bell pairs at $n=L/2$ and the other intermediary states. In the case of the $y$-initial state (b, d), a drop in $E_g$ is seen at $n=L,L+1$. This is due to the system reaching states with high number of particles involved in the entanglement (the GHZ states).}
\label{n10Edist}
\end{figure}

\begin{figure}[t]	
\centering
\includegraphics[angle=0,width=\linewidth]{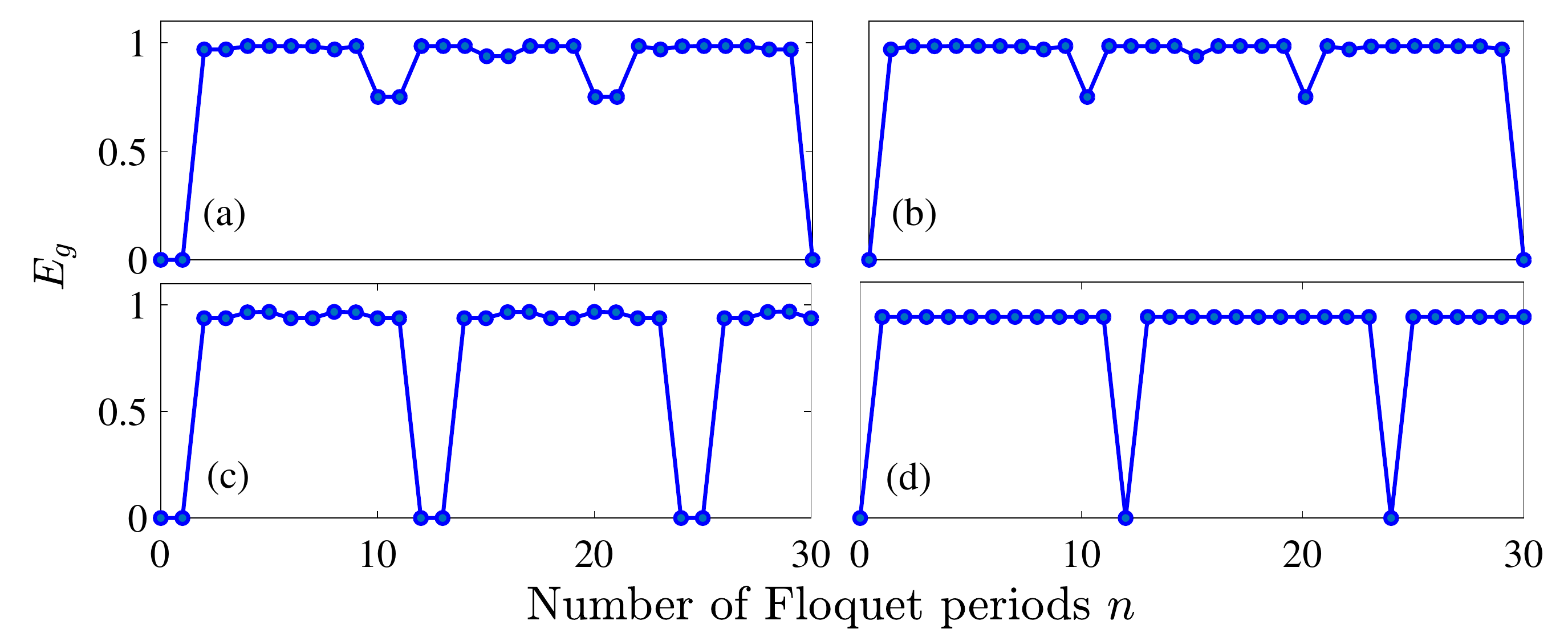}
\caption{Variation of the geometric measure of entanglement  $E_g$ with consecutive Floquet periods in the $\mathcal{U}_x$ system of sizes $L=10$ (a, b) and $L=8$ (c, d) starting from $z$-initial state (a, c) and $y$-initial state (b, d). For system size $L=10$, the the $E_g$ values are high at most points. The points with low values of $E_g$ ($n=10,11,...$) are points where there are high number of particles involved in the entanglement.
A similar plot is seen for other even system sizes (upto $L\leq12$) except for $L=8$. For $L=8$ (c, d) the plot is similar to that seen in the $\mathcal{U}_0$ system (\cref{n10Edist}(a, c))}
\label{n10xEdist}
\end{figure}

\begin{figure}[t!]
\centering
\includegraphics[angle=0,width=\linewidth]{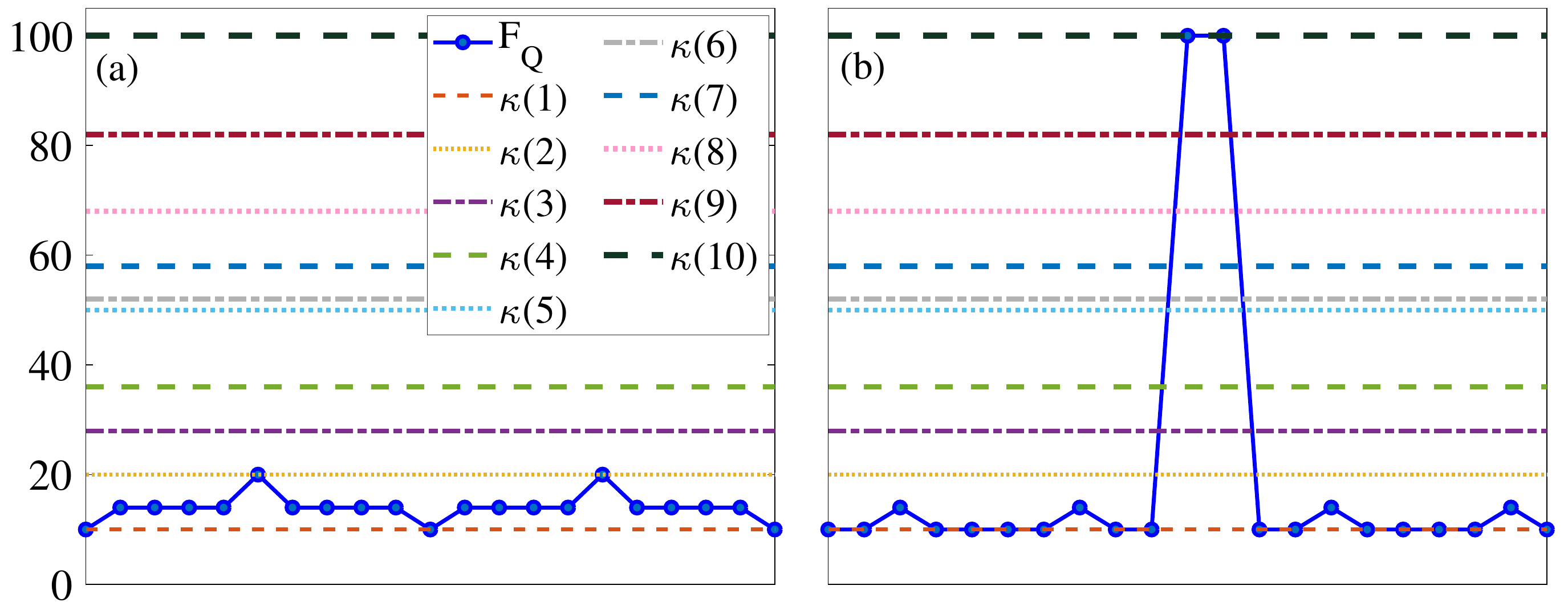}
\includegraphics[angle=0,width=\linewidth]{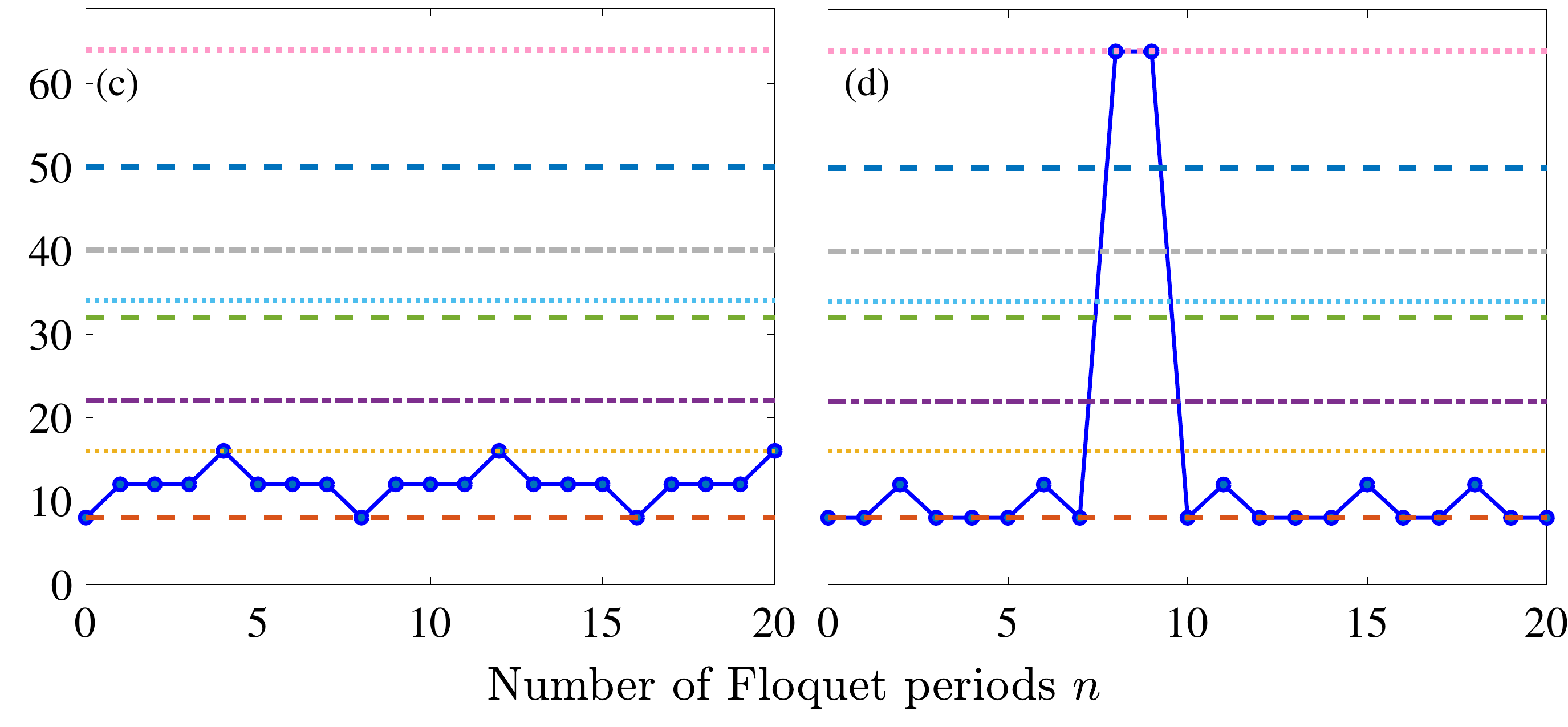}
\caption{Variation of the maximum value of $F_Q$ with consecutive Floquet periods in the $\mathcal{U}_0$ system of sizes $L=10$ (a, b) and $L=8$ (c, d) starting from $z$-initial state (a, c) and $y$-initial state (b, d). $\kappa(k)$ is the maximum value of QFI for $k$-producible states. A value of $F_Q$ greater than $\kappa(k)$ indicates that the state has at least $k+1$-particle entanglement. In (a) and (c), the system reaches states with two particle entanglement, while in (b) and (d) system reaches states with $L$ particle entanglement.}
\label{n10FQ}
\end{figure}

\begin{figure}[t!]
\centering
\includegraphics[angle=0,width=\linewidth]{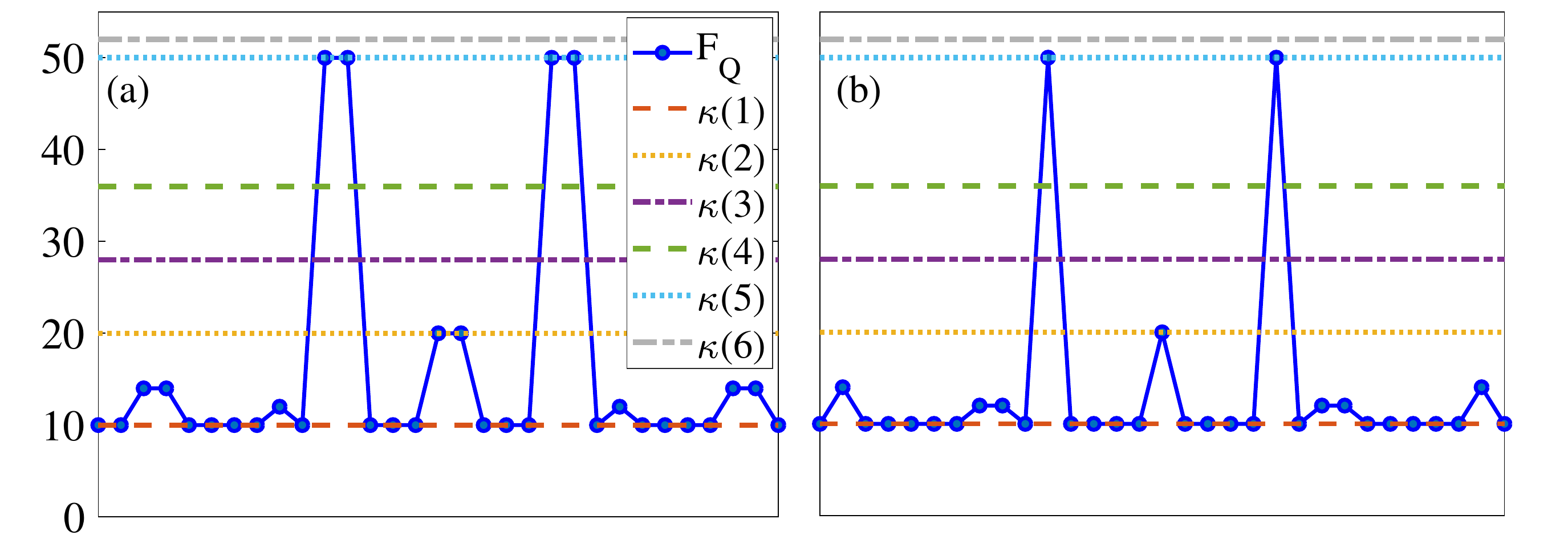}
\includegraphics[angle=0,width=\linewidth]{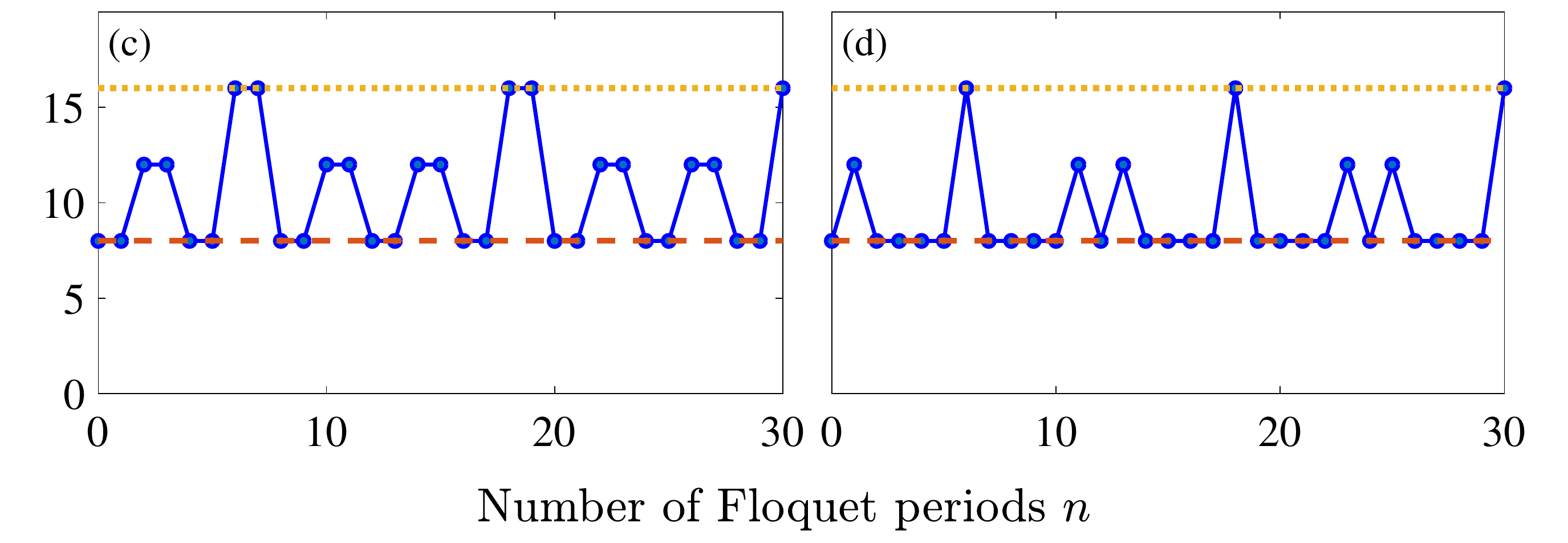}
\caption{Variation of the maximum value of $F_Q$ with consecutive Floquet periods in the $\mathcal{U}_x$ system of sizes $L=10$ (a, b) and $L=8$ (c, d) starting from $z$-initial state (a, c) and $y$-initial state (b, d). $\kappa(k)$ is the maximum value of QFI for $k$-producible states. The system with even system size is seen to have at least $L/2$ particle entanglement (checked for $L\leq 12$). However, $L=8$ (c,d) is an exception as the $F_Q$ plot suggests only at least two particle entanglement.}
\label{n10xFQ}
\end{figure}

A more well known measure of entanglement in the complete system is the geometric measure of entanglement \cite{Wei2003a} (also known as the distance measure of entanglement) and is given by:
\begin{equation}
E_g=1-\Lambda^2,
\label{E_g}
\end{equation}
where $\Lambda = \max_{\Phi} |\langle \psi | \Phi \rangle |$, with the maximization done over all the possible separable product states $|\Phi\rangle=\otimes_{i=1}^L |\phi_i\rangle$. 
The $\mathcal{U}_0$ system with $z$-initial state after $L/2$ kicks gets to the state with $L/2$ Bell pairs which is expected to have high entanglement. However, the plots of the geometric measure of entanglement in \cref{n10Edist}(a, c) show that all the intermediate states (from $n=1$ to $n=L-1$) of the system have high entanglement. Also from the other plots in \cref{n10Edist,n10xEdist}, we can see that even in the other cases, the system has states with high entanglement. 
At certain points \big(such as $n=10,11$ in \cref{n10Edist}(b) and \cref{n10xEdist}(a)\big), the system has smaller values of $E_g$ and AEE (refer \cref{n10En}). Further in the paper, we have shown that these are points where the system has a high number of particles involved in the entanglement \big(refer \cref{n10FQ}(b) and \cref{n10xFQ}(a)\big).

This suggests a variation of the principle of monogamy of entanglement, where we would hypothesise that the higher bound on the value of entanglement measures such as $E_g$ and AEE for a multipartite quantum state with $n$ particles involved in the entanglement decreases as $n$ increases. A simple example where this is evident is a system of four spins with states 
\begin{eqnarray}
 \st{\phi_1} &=& \left(\frac{\st{00}+\st{11}}{\sqrt{2}}\right) \otimes \left(\frac{\st{00}+\st{11}}{\sqrt{2}} \right) \text{ and} \nonumber \\ 
\st{\phi_2} &=& \left(\frac{\st{0000}+\st{1111}}{\sqrt{2}}\right) \end{eqnarray}
The state with four particles involved in the entanglement $\phi_1$ (the four particle GHZ state shown in \cite{Wei2003a} to have $\Lambda=1/\sqrt{2}$ and $E_g=1/2$ ) has a lower value of $E_g$ than the state with two particles involved in the entanglement $\phi_2$ (the product of two Bell pairs with $\Lambda=1/2$ and $E_g=3/4$). 

To investigate the number of particles involved in the entanglement seen in these systems, we measure the quantum Fisher information (QFI) of the states of the system. The QFI of a pure state $\st{\psi}$ associated with a linear observable $\hat{O} =\frac{1}{2} \sum\limits_{i=1}^L \hat{n}_i .\vec{\sigma}_i$ (where $\hat{n}_i$ for $1 \leq i \leq L$ are a unit vectors) \cite{Toth2012,Hyllus2012,Hauke2016,Pappalardi2017} is given by
\begin{equation}
F_Q(\hat{O})= 4 \langle \Delta \hat{O} \rangle^2 ,
\label{F_QO}
\end{equation}
where $\langle \Delta \hat{O} \rangle^2= \langle \psi | \hat{O}^2 | \psi \rangle - \langle \psi | \hat{O} | \psi \rangle^2  $. If for some linear observable $\hat{O}$, the inequality 
\begin{equation}
F_Q \leq  \floor*{ \frac{L}{k} } k^2+\left( L-\floor*{\frac{L}{k}} k \right)^2 
\label{F_Q}
\end{equation} 
is violated, then the state $\st{\psi}$ has at least $k+1$-partite entanglement (here $\floor{x}$ denotes greatest integer lesser than or equal to $x$). To get an estimate of the number of particles involved in the entanglement of a particular state, we maximize $F_Q$ over the space of linear observables $\hat{O}$ (parametrized by the unit vector $\hat{n}$) and then find the largest value of $k$ that violates the inequality (\ref{F_Q}).

In the $\mathcal{U}_0$ system, with the $z$-initial state, we see that states of the system after $n$ Floquet periods for $2 \leq n \leq L-1$ all involve at least two particle entanglement (refer \cref{n10FQ}(a, c)). A peak in the QFI is seen at $n=L/2$ when the system reaches the state of product of $L/2$ Bell pairs. 
The $\mathcal{U}_0$ system with the initial state of spins aligned along $y$-direction after $L$ and $L+1$ kicks gets to the $L$ particle GHZ (Greenberger-Horne-Zeilinger) state in the $y$ and $x$ direction respectively. Hence, we see peaks suggesting $L$ particle entanglement in the QFI plots of this system (\cref{n10FQ}(b, d)).
For system size $L=8$, the plots of QFI (\cref{n10xFQ}(c, d)) for the $\mathcal{U}_x$ system indicate that the system has states with at least two particle entanglement. For other even system sizes, the QFI plots (\cref{n10xFQ}(a, b)) suggest that the system has states with at least $L/2$ particle entanglement (checked upto $L=12$).
These are states that have high genuine multipartite entanglement and cannot be expressed in simple forms in the standard basis.

The reference \cite{Hyllus2012} states that equality in \cref{F_Q} is only possible if the state is a product state of $\floor*{ \frac{L}{k} }$ $k$-particle GHZ states and a $\left( L-\floor*{\frac{L}{k}} k \right)$ particle GHZ state. However, our numerical calculations shows many points on the QFI plots that represent states satisfying the equality in \cref{F_Q} but are not a product of GHZ states. A simple example of a non-GHZ state satisfying the equality in \cref{F_Q} is the state $\mathcal{U}_x \st{\psi_o}$, where $\st{\psi_o}$ is a product of two GHZ states:
\begin{eqnarray}
 \st{\psi_o} &=& \left(\frac{\st{0\cdots0}_{1\cdots \frac{L}{2}}+\st{1\cdots 1}_{1\cdots \frac{L}{2}}}{\sqrt{2}} \right) \otimes \nonumber \\
& &\left( \frac{\st{0\cdots0}_{\frac{L}{2}+1 \cdots L}+\st{1\cdots 1}_{\frac{L}{2}+1\cdots L}}{\sqrt{2}} \right). 
\end{eqnarray} 
We know that the state $\st{\psi_o}$ has $F_Q=2L$ and satisfies equality of \cref{F_Q} for $k=2$. The state $\mathcal{U}_x \st{\psi_o}$ can be numerically verified to have maximum QFI of $F_Q=2L$ but this state cannot be expressed as a product of two GHZ states in any basis. This can be verified by calculating the entropy of all possible partitions. The state $\mathcal{U}_x \st{\psi_o}$ has no partition with zero entropy which implies that it cannot be expressed as a product state in any basis. For $L=4$, 
\begin{eqnarray}
 \mathcal{U}_x \st{\psi_o} &=& \left(\frac{\st{00}+\st{11}}{\sqrt{2}}\right) \otimes \left(\frac{\st{00}+\st{11}}{\sqrt{2}} \right) \nonumber \\ 
 & &+ i \left(\frac{\st{01}+\st{10}}{\sqrt{2}} \right) \otimes \left( \frac{\st{01}+\st{10}}{\sqrt{2}}\right).
\end{eqnarray}

We have considered initial product states polarized along the $x$,$y$ or the $z$ direction for our study. However, any general initial state may be considered and a similar study may be carried out to identify high multiparty entangled states obtained by time evolution using $\mathcal{U}_x$ and $\mathcal{U}_0$ operators. Since the system is periodic in time, only a finite number of states is obtained by the Floquet time evolution for a given initial state.
A similar study can be done for odd systems sizes and also periodic chains (refer \cref{summary_table}). The results in these cases cannot be directly extended from the results obtained above.

\section{Conclusion}
\label{conclusion}
We have considered two Floquet systems with Floquet operators $\mathcal{U}_0$ \big[\cref{U_0}\big] and $\mathcal{U}_x$ \big[\cref{U_x}\big] and initial states with all spins polarized in a specific direction.
We have shown that these systems are periodic in time as explained by the quasi-energies of these systems that have degeneracies and a constant gap. The quasi-energies of the $\mathcal{U}_0$ system, which is integrable, are seen to be of the form $n\pi/(2L)$ (where $L$ is the system size and $n$ is an integer such that $-2L<n\leq 2L$). We then evolved simple product initial states and analyzed the average entanglement entropy and the geometric measure of entanglement of the time evolved states to show that they have high multiparty entanglement. We have also calculated the quantum Fisher information of these states to identify those that have a high number of parties involved in the entanglement.

We have shown that many states with high genuine multiparty entanglement can be obtained by the time evolution of simple product initial states.
The $\mathcal{U}_0$ system with the initial state of spins aligned along the $y$-direction ($x$-direction) generates $L$-particle GHZ states after $L$ and $L+1$ ($L-1$ and $L$) Floquet periods. The $\mathcal{U}_x$ system with the initial state of all spins aligned along one of the three directions ($x$, $y$ or $z$-direction) reaches a state with at least $L/2$ particle entanglement for most system sizes. A summary of the multipartite entanglement generated for different initial states and boundary conditions have been outlined in \cref{summary_table}.

We propose that these Floquet systems, which have only nearest neighbor interactions, can be used to generate states with high multiparty entanglement. Recent experiments have shown that Floquet spin systems can be realized by systems such as trapped ions \cite{Zhang2017} and nitrogen-vacancy (NV) spin impurities in diamond \cite{Choi2017}. These systems can potentially be used for the physical realization of the considered Floquet systems to generate high multiparty entangled states which can further be used as a resource for quantum computation, quantum cryptography and the quantum internet \cite{McCutcheon2016,Chang2016}.

\section{Acknowledgement}
SKM would like to thank Arul Lakshminarayan and GKN would like to thank Vadim Oganesyan and Sarang Gopalakrishnan for fruitful discussions and comments on the work.
RS acknowledges support from Department of Science and Technology, India through the Ramanujan Fellowship.

\bibliographystyle{apsrev4-1}
\bibliography{Multiparty_Entanglement}

\appendix

\section{Summary Tables}
\label{summary_table}
The following tables give the summary of the multiparty entanglement seen in the $\mathcal{U}_x$ and $\mathcal{U}_0$ systems in the different cases of boundary conditions and system sizes that have been studied. Here, $n$ denotes the number of Floquet periods after which the system is considered.

\begin{widetext}

\begin{table}[h!]

\textbf{For even system sizes with open boundary conditions:}
\begin{tabular}{|c|l|l|}
\hline
\textbf{Initial State} & \multicolumn{1}{c|}{\textbf{$\mathcal{U}_0$ system}} & \multicolumn{1}{c|}{\textbf{$\mathcal{U}_x$ system}} \\ \hline
$z$ & \begin{tabular}[c]{@{}l@{}}$L/2$ Bell states at $n=L/2$.\\ Entanglement structure periodic with period $2L$. \\ System periodic with period $4L$.\end{tabular} & \begin{tabular}[c]{@{}l@{}}$L/2$ particle entanglement at certain values of $n$ with\\ an exception for $L=8$ (checked up to $L=10$). \\ System periodic with period not simple function of $L$.\end{tabular} \\ \hline
$y$ & \begin{tabular}[c]{@{}l@{}}$L$ particle entanglement at $n=L$ (GHZ state in \\ y-basis) and $n=L+1$ (GHZ state in $x$ basis). \\ Entanglement structure periodic with period $2L$. \\ System periodic with period $4L$.\end{tabular} & \begin{tabular}[c]{@{}l@{}}$L/2$ particle entanglement at certain values of $n$ with\\ an exception for $L=8$ (checked up to $L=10$). \\ System periodic with period not simple function of $L$.\end{tabular} \\ \hline
$x$ & Same as the case of $y$-initial state with $n'=n-1$. & Same as the case of $z$-initial state with $n'=n-1$. \\ \hline

\end{tabular}
\end{table}

\begin{table}[h!]

\textbf{For odd system sizes with open boundary conditions:}
\centering
\begin{tabular}{|c|l|l|}
\hline
\textbf{Initial State} & \multicolumn{1}{c|}{\textbf{$\mathcal{U}_0$ system}} & \multicolumn{1}{c|}{\textbf{$\mathcal{U}_x$ system}} \\ \hline
$z$ & \begin{tabular}[c]{@{}l@{}}3 particle entanglement observed till $L=9$.\\ Entanglement structure periodic with period $L$.\\ System periodic with period $4L$.\end{tabular} & \begin{tabular}[c]{@{}l@{}}$L$ particle entanglement at certain values of $n$ \\ with a different entanglement profile \\ (except for $L=7$)(checked up to $L=7$). \\ System periodic with period not simple function of $L$.\end{tabular} \\ \hline
$y$ & \begin{tabular}[c]{@{}l@{}}$L$ particle entanglement at $n=L$ (GHZ state \\ in y-basis) and $n=L+1$ (GHZ state in $x$-basis). \\ Entanglement structure periodic with period $2L$.\\ System periodic with period $4L$.\end{tabular} & \begin{tabular}[c]{@{}l@{}}$L$ particle entanglement at certain values of n \\ (except for $L=7$)(checked up to $L=7$). \\ System periodic with period not simple function of $L$.\end{tabular} \\ \hline
$x$ & Same as the case of $y$-initial state with $n'=n-1$. & Same as the case of $z$-initial state with $n'=n-1$. \\ \hline
\end{tabular}\\
%
%
\vspace{4em}
\textbf{For even system sizes with closed boundary conditions:}
\centering
\begin{tabular}{|c|l|l|}
\hline
\textbf{Initial State} & \multicolumn{1}{c|}{\textbf{$\mathcal{U}_0$ system}} & \multicolumn{1}{c|}{\textbf{$\mathcal{U}_x$ system}} \\ \hline
$z$ & \begin{tabular}[c]{@{}l@{}}2 particle entanglement in system sizes $L=4n$ and no\\ entanglement in $L=4n+2$ (as the FQ plot suggests).\\ System periodic with period $L$.\end{tabular} & \begin{tabular}[c]{@{}l@{}}$L/2$ particle entanglement at certain values of $n$ with\\ an exception for $L=8$ (checked up to $L=10$). \\ System periodic with period not simple function of $L$.\end{tabular} \\ \hline
$y$ & \begin{tabular}[c]{@{}l@{}}$L/2$ particle entanglement at $n=L/2$ and $n=L/2+1$\\  (superposition of ferromagnetic and anti-ferromagnetic \\ GHZ state in y-basis and $x$-basis respectively). \\ Entanglement structure periodic with period $L$.\\ System periodic with period $L$.\end{tabular} & \begin{tabular}[c]{@{}l@{}}$L/2$ particle entanglement at certain values of $n$ with\\ an exception for $L=8$ (checked up to $L=10$). \\ System periodic with period not simple function of $L$.\end{tabular} \\ \hline
$x$ & Same as the case of $y$-initial state with $n'=n-1$. & Same as the case of $z$-initial state with $n'=n-1$. \\ \hline
\end{tabular}
\end{table}

\end{widetext}

\end{document}